\documentclass[epj,referee]{svjour}
\usepackage{amsmath,amssymb}
\usepackage{graphicx}
\usepackage[dvips]{hyperref}
\begin{document}
\title{Chaos in free electron laser oscillators}

\author{C. Bruni\inst{1}\thanks{\emph{corresponding author:}
    bruni@lal.in2p3.fr}  \and R. Bachelard\inst{2} \and D. Garzella\inst{3} \and G. L. Orlandi\inst{4}  \and M. E. Couprie\inst{2} }
%
%
\institute{LAL, Universit\'e Paris-Sud 11, UMR 8607, Bâtiment 200, 91898 Orsay cedex \and Synchrotron SOLEIL, L'Orme des Merisiers, Saint-Aubin - BP 48, 91192 Gif-sur-Yvette Cedex, France\and CEA/DSM/DRECAM/SPAM, B\^{a}t. 522, 91191 Gif-sur-Yvette, France \and ENEA, C.R. Frascati FIM-FISACC, Via E. Fermi 45, 00044 Frascati, Italy }
\date{Received: \today / Revised version: date}
%
\abstract{The chaotic nature of a storage-ring Free Electron Laser (FEL) is investigated. The derivation of a low embedding dimension for the dynamics allows the low-dimensionality of this complex system to be observed, whereas its unpredictability is demonstrated, in some ranges of parameters, by a positive Lyapounov exponent. The route to chaos is then explored by tuning a single control parameter, and a period-doubling cascade is evidenced, as well as intermittence.}


\maketitle
\section{Introduction}

Perturbations of systems have been shown to generically lead to bifurcations in the dynamics, for example in biology \cite{Nature98Coffey}, ecology \cite{Nature93Stone}, chemistry \cite{Nature93Petrov} or astrophysics \cite{Nature92Milani}. The bifurcations is eventually replaced by chaos, which has been observed in a wide range of plasma/wave interaction such as plasma systems \cite{PLA00Hur} and FELs \cite{PRE93Hahn,PRL03Li}. The route to chaos frequently occurs in these systems, and links have been done between FEL chaos and plasma chaos \cite{APL98lee,IEEE00Lee}.

In the case of FELs, a thorough understanding of the nonlinear dynamics of the system is of paramount importance, since their ultimate goal is to deliver stable light pulses for e.g. chemistry, biology and surface studies experiments \cite{NIMA01Renault,RSI94Couprie,PRB00marsi}. In particular, the Sensitivity to Initial Conditions (SIC), one of the manifestation of chaos, can drive the system into very different dynamical regimes, even for tiny variations of the initial state. Storage Ring FEL (SRFEL) are particularly rich from a nonlinear dynamics point of view. The evolution of the pulse train shape from stationary states to limit cycles \cite{PRL92Billardon} has been demonstrated. The transition from a stationary state to a limit cycle arises through a Hopf bifurcation \cite{PRL04Deninno}. Using control of chaos methods, these limit cycle regimes were suppressed by stabilizing the coexisting unstable stationary state \cite{PREBielawski04,PRL04Deninno,EPJD06bruni}. For example, the ELETTRA FEL needs this type of control because of the lack of natural stationary state. A topological approach has also being investigated and stable orbits have been identified \cite{EPJD07bachelard}. FEL nonlinear studies have also enabled to highlight intensity holes in the spectrum associated to drift instabilities \cite{PRL05Bielawski} in the spectro-temporal diagram of the FEL intensity distribution. More generally, a SRFEL is a system which presents a localized structure in presence of advection. As a consequence, this phenomenon observed in SRFEL systems is expected to affect a range of systems displaying localized solutions with either local or global coupling. The drift instabilities can be stabilised thanks to an optical feedback inspired from the community of conventional laser \cite{OC90Beaud,OL90New}, a technique already tested on the UVSOR FEL \cite{SergePrivate}. 

Here, we are particularly interested in the chaotic features of the dynamics and their conditions of emergence. Chaos studies have first been performed on the Super-ACO and ACO storage ring FEL thanks to an external modulation \cite{PRL90Billardon} and explained with a three-dimensional model. It was shown that the recorded temporal series were chaotic and assumption was made that, as for conventional lasers, the route to chaos goes through a period doubling cascade \cite{PRE95WenJie}. Afterward, a more complete model, which integrated the presence of an asynchronism between light and electron pulses (called detuning), was proposed \cite{EPJD03Deninno} and employed to study the presence of chaos in the dynamics. It revealed that a period-doubling bifurcation can arise under changes in two control parameters.

We here characterize this chaotic dynamics by deriving the embedding dimension and the Lyapunov exponent of the dynamics. Then, a single control parameter is finely tuned to explore the route to chaos, and to reveal the period-doubling cascade, as well as the presence of intermittence.

\section{FEL temporal dynamics}



During the propagation through a periodic permanent magnetic field structure (an undulator), a
relativistic electron beam emits an electromagnetic wave at its fundamental wavelength and its odd
harmonics \cite{JP83Elleaume}. The light amplification results from the interaction between the
relativistic electron beam in the undulator and the store optical pulse in the optical cavity. The laser gain \cite{JAP71Madey} depends both on the electron beam and the undulator characteristics. Amplification takes place at a wavelength close to the fundamental, tunable by a modification of the magnetic field of the undulator. The gain reaches the saturation level defined by cavity total losses at the expense of an increase of the energy spread of the electron bunch.

The laser is pulsed at the repetition frequency of the electron beam in the undulator (in the MHz range), and the length of the optical resonator must be carefully adjusted to satisfy the longitudinal synchronization between the optical pulses and the electron bunches. In practice, on storage rings FEL, the synchronization condition is controlled by the frequency of the Radio-Frequency (RF) cavity, since it modifies the orbit length of the electrons. The temporal detuning $\delta$ is the difference between the round trip period of the light pulse in the optical cavity $\tau_{R}$ and the electron bunch spacing $T_{0}$. A perfect synchronism is required to obtain a stable stationary state of the laser (pulsed at the MHz range) \cite{PRL05Bielawski}, for which the pulse train shape is constant. Nevertheless, beyond a given threshold of detuning, the laser exhibits a pulsed shape in the millisecond range. This can be simply understood with the model describing the coupled evolution of the laser pulse intensity $I$ (pulse train shape) and of the normalized energy spread $\Sigma$ of the electron bunch \cite{JP84Elleaume}:
\begin{subequations}
\begin{eqnarray}
\frac{dI}{dt}&=&\frac{I}{T_{0}}\left(G_0-L\right)\left(1-\Sigma\right)+i_{s} \label{eq:ElleaumeIa}
\\  \frac{d\Sigma}{dt}&=&\frac{-2}{\tau_{s}}\left(\Sigma-I\right) \label{eq:ElleaumeIb}
\end{eqnarray}
\end{subequations}
with $\Sigma=(\sigma^{2}-\sigma_{0}^{2})/(\sigma_{e}^{2}-\sigma_{0}^{2})$, $\sigma$ being the energy spread,
$\sigma_{0}$ the energy spread at the laser start up, $\sigma_{e}$ the energy spread at the laser asymptotic equilibrium
state. $T_{0}$ is the spatial period between two electron bunches, $G_0$ the maximum laser gain (which can be analytically determined \cite{JP83Elleaume,Preprint77Vinokurov} and depends on the characteristics of the undulator and electron beam) and $L$ the cavity losses. The synchrotron damping time $\tau_{s}$ is related to the electron bunch relaxation dynamics. The spontaneous emission $i_{s}$ represents the initial noise from which the laser starts. The equilibrium solution of the equations (\ref{eq:ElleaumeIa}) and (\ref{eq:ElleaumeIb})
corresponds to $I=1$ and $\Sigma=1$. For a small perturbation around $I$ and $\Sigma$ and by neglecting the noise, the equation 1 can be linearised, leading to a pulsed laser train shape at a resonant frequency:
\begin{equation}\label{eq:Fr}
  F_{r}=\frac{1}{\pi}\sqrt{\frac{G_{0}-L}{2T_{0}\tau_{s}}}.
\end{equation}

The model given by equations (\ref{eq:ElleaumeIa}) and (\ref{eq:ElleaumeIb}) reproduces well the presence of a constant and a pulsed pulse train shape of the laser. Yet it considers only the pulse train shape, and does not take into account the temporal mismatch between the light and electron pulses, the so-called detuning $\delta$. To modelize this detuning, a longitudinal distribution for the intensity has to be introduced, which describes the pass-to-pass interaction process \cite{EPJD03Deninno}:
\begin{subequations}\label{modelEPJD}
\begin{eqnarray}
  y_{n}(\tau)&=&(1-L)\left[1+G_{n}(\tau)\right]y_{n-1}(\tau)+i_{s} \exp\left(\frac{-\tau_{n}^2}{2\sigma_{\tau_n}^2}\right),\label{eq:LAS1a}
\\ \Sigma_{n}&=&\Sigma_{n-1}-\frac{2T_{0}}{\tau_{s}}[\Sigma_{n-1}-\int_{-\infty}^{+\infty}{y_{n-1}(\tau)}d\tau],\label{eq:LAS1b}
\end{eqnarray}
\end{subequations}
where $y_{n}$ stands for the longitudinal distribution (centered on the center of mass of the electron bunch) of the laser pulse at the $n$-th pass in the gain medium (electron bunch in the undulator). This bunch has a normalized energy spread $\Sigma$ at the $n$-th pass and a rms longitudinal width $\sigma_{\tau}$, supposed to be proportional to the energy spread as conventionally used in the accelerator community.

These equations include the effect of the detuning by adding a cumulative delay between the electron bunches and
the laser pulse:
\begin{equation}\label{eq:delay}
  \tau_{n+1}=\tau_{n}+\delta.
\end{equation}
This delay modifies the longitudinal overlap between the Gaussian electron bunch distribution and the laser pulse, which reduces the gain as:
\begin{equation}\label{eq:gain}
  G_{n}(\tau)=G_{0}\left[\frac{L}{G_{0}}\right]^{\Sigma_{n}}\exp\left({\frac{-(\tau_{n}+\delta)^{2}}{2\sigma_{\tau}^{2}}}\right).
\end{equation}
The laser intensity at pass $n$ can be expressed as the zero-order momentum of the laser intensity distribution :
\begin{equation}\label{eq:I0}
I_{n}=\int_{-\infty}^{+\infty} y_{n}(\tau)d\tau.
\end{equation}

\begin{figure}
  \centering
  \includegraphics{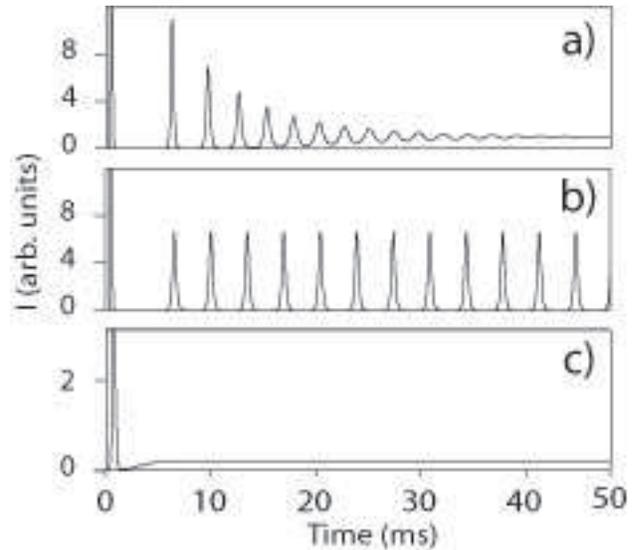}
\caption{Numerical simulations of the laser intensity versus time for different
detuning values: a) $\delta=0$ $fs$, b) $\delta=6$ $fs$, c) $\delta=36$ $fs$. Simulations parameters (corresponding to the Super-ACO FEL):
$\tau_{R}=120~ns$, $G_{0}=1.5~\%$, $L=1~\%$, $\sigma_{0}=6.5$x$10^{-4}$, $\sigma_{e}=9.3$x$10^{-4}$ ,
$\sigma_{\tau}=107~ps$, $\tau_{s}=9~ms$, $i_{s}=~10^{-9}$ }\label{fig:time-struc-LAS-det}
\end{figure}

At perfect tuning, the laser tends towards an asymptotic solution (constant pulse train shape), characterized by $I=1$ and $\Sigma=1$ (see figure \ref{fig:time-struc-LAS-det}a), in agreement with model (\ref{eq:ElleaumeIa},\ref{eq:ElleaumeIb}). When increasing the detuning, the laser intensity first remains constant in the stationary regime, then beyond a given threshold, it becomes pulsed (pulsed train shape) at a few hundred Hertz (see figure \ref{fig:time-struc-LAS-det}b). At this point, the resonant frequency $F_{R}$ changes with the detuning \cite{NIMA02DeNinno}. If one keeps increasing $\delta$, a second threshold is crossed and the system quickly reaches a stationary state where both $I$ and $\Sigma$  are constant, and less than unity (see figure \ref{fig:time-struc-LAS-det}c). It has been shown \cite{EPJD06bruni} that this model reproduces well the experimental dynamics of different FEL facilities \cite{PAC01Wang,NIMA03DeNinno,NIMA02Yamada,PRE98Roux,NIMA00Hosaka}. For example, the absence of experimental stationary state at perfect tuning on the ELETTRA FEL has been explained by the small width of the central cw zone \cite{EPAC04Bruni}. Note that similar dynamics have been observed on conventional mode-locked lasers, which have been demonstrated to be governed by analogous equations \cite{JPIV06Bruni}.

In the following, the properties of the system are investigated when the detuning is modulated at a frequency close to the resonant frequency of the system. Experimentally, the laser is first adjusted at perfect tuning (with a 0.1 fs precision), then a modulation $\delta (t)$ is introduced with a function generator, which modulates the RF frequency at 100 MHz\footnote{Storage ring FELs are based on the emission of synchrotron radiation by charged particles in magnetic devices (such as dipoles or undulators), which consists in a conversion of kinetic energy into a radiation. To compensate for the loss of energy at each interaction, and so at each pass in the cavity, the RF boosts the electrons: Its frequency has to be linked to the revolution period of the electrons in the ring.}:
\begin{equation}\label{eq:modulation}
  \delta(t)=a\sin(2\pi ft),
\end{equation} 
with $f=320~Hz$. The control parameter $a$ is tuned manually by adjusting the amplitude of the function
generator signal. During the experiment, the laser
intensity and its temporal spectrum (obtained by Fourier Transform oscilloscope routine) are simultaneously recorded allowing changes in the regimes of the system to be detected.

\section{Characterization of the chaos}

\begin{figure}
        \begin{center}
            \includegraphics[scale=0.7]{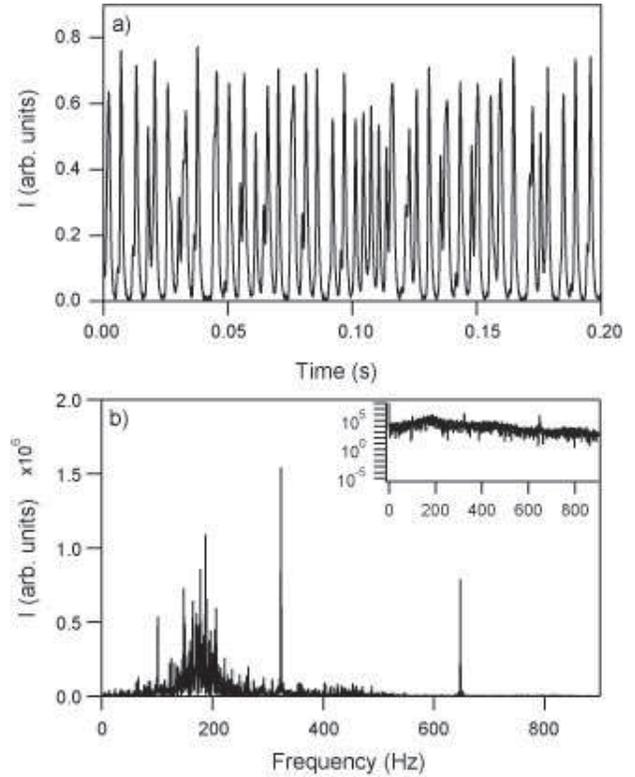}
            \caption{Super-ACO chaotic dynamics for a detuning modulation of $a=16.5fs$, at frequency $320~Hz$. a) Laser intensity vs. time and its, b) Fourier Transform,
              with logarithm scale
              representation in the inset. The intensity measurements were performed with a photomultiplier (Hamamatsu R928), and analyzed by an oscilloscope (Lecroy LT264M, $350~MHz$, $10^9$ samples/s) triggered
by the trigger signal of the function generator (Stanford Research System DS345). The acquisition was set to 25000 samples/s on a two-second time scale. The time series comprise at least 560 periods.} \label{Exemple-2T-chaos}
        \end{center}
      \end{figure}

In this section, the chaotic nature of the experimental FEL dynamics is explored, under a modulation of detuning. Analysis of the nonlinearities are performed on the experimental signals. Figure
\ref{Exemple-2T-chaos} displays Super-ACO SRFEL macropulses, which have no apparent regularity and whose intensity peaks are strongly fluctuating. This is confirmed by inspecting the spectrum, which presents a continuous background between $0$ and $640~Hz$ (see figure \ref{Exemple-2T-chaos}b), which characterizes the aperiodicity of the time series. Nonetheless, one can note that the spectrum exhibits a peak at twice the modulation frequency.

In order to get a deeper insight into the dynamics, the resulting aperiodic time series are analyzed with the false nearest neighbours method to estimate the embedding dimension and with the calculation of the maximum Lyapunov exponent to appreciate the sensitivity to initial conditions \cite{RMP85Eckman}.

\begin{figure}
       \begin{center}
            \includegraphics[scale=0.75]{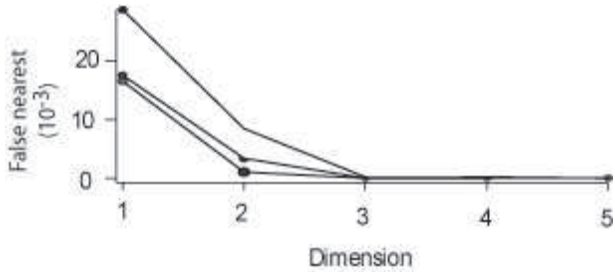}
            \caption{Ratio of false nearest neighbours versus the phase space dimension for
              three experimental time series of the laser intensity
              $I(t)$ for a modulation frequency $f=320~Hz$ and a=33 fs or 51 fs, calculated with the TISEAN programs
\cite{tisean1,tisean2,PRE94Kantz}, using the delay coordinates.} \label{Faux-voisins}
        \end{center}
    \end{figure}


Generically, introducing additional degrees of freedom into a system generates more complex dynamical regimes, since it can explore a phase space of larger dimension. Yet, systems with many degrees of freedom can exhibit macrostructures -both temporal or spatial- of a surprising simplicity. In order to determine the $\it effective$ complexity of a dynamics, one can resort, for example, to the $\it embedding~dimension$, which gives an estimate of the number of variables necessary to describe the dynamics. Such a reduction of the complexity is crucial, since in our case a continuous description of the light pulse distribution necessarily implies an infinite-dimensional phase space. The minimal dimension of the space in which a time series evolves can be estimated by the false nearest neighbours method \cite{PRA92Kennel}. One point $I(t_{i})$ is ``nearest'' to another one $I(t_{i+1})$ if the distance $R_{i}$ between the two points is not higher than a heuristic threshold $R_{t}$. Once the number of points for which $R_{i}<R_{t}$ is zero or close to zero, then the embedded dimension is reached.

Figure \ref{Faux-voisins} relates the ratio of false nearest neighbours for three aperiodic time series of the laser intensity from Super-ACO experiment. This ratio falls to zero for a dimension $d=3$, which tends to indicate that the dynamics can be embedded into a three-dimensional space. Despite of the complexity of the system, this result suggests that it may be possible to describe the dynamics in a low-dimensional model. 


\begin{figure}
       \begin{center}
            \includegraphics[scale=0.8]{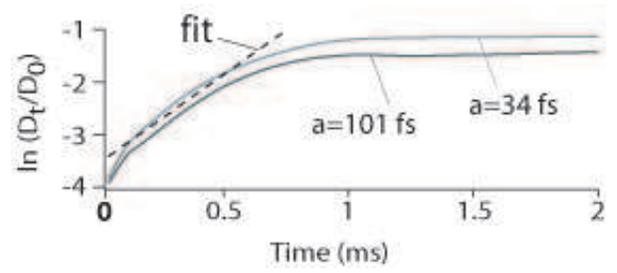}
            \caption{Calculation of
              $\ln(\frac{\Delta_{t}}{\Delta_{0}})$ as a function of
              time for different time series for a modulation
              frequency $f=320~Hz$. The fit has been realized on the
              linear part. Programs of TISEAN
\cite{tisean1,tisean2} has been used to draw this plot, using the delay coordinates.} \label{Lyapunov}
        \end{center}
      \end{figure}

The sensitivity to initial conditions can be characterized by its Lyapunov exponent, that is, the divergence rate in phase space of two close trajectories \cite{RMP85Eckman,PRA76Benettin}. Even though only the intensity can be measured, some techniques have been developed to rebuild a phase space from a single variable, such as the delay coordinates [$I(t)$, $I(t+\delta\tau)$, $I(t+2\delta\tau)$,..., $I(t+m\delta\tau)$], where $\delta\tau$ corresponds to the delay.  Although it provides a limited access to the dynamics, it is assumed that it can give a representative estimation of the chaotic properties of the system. 

Let us consider two close initial conditions distant by $\Delta_0$ in the rebuilt phase space. If we call $\Delta_t$ the distance between the states at $t_1+t$ and $t_2+t$, the Lyapunov exponent $\lambda_{l}$ is defined as:
\begin{equation}\label{lyapunov}
 \lambda_l=\lim_{t\rightarrow \infty} \lim_{\Delta_0\rightarrow 0} \log{\frac{\Delta_t}{\Delta_0}}
\end{equation}
The Lyapunov exponent is then calculated in the phase space associated to the delay coordinates, with $m=2$ in agreement with the result of the false nearest neighbours.

Figure \ref{Lyapunov} represents the exponential growth in time of nearby
trajectories. The linear increase relates the divergence, between $0.1$ and $0.5~ms$, where the linear fit allows
the maximum Lyapunov exponent between $3.2$ and $3.8~ms^{-1}$ to be calculated. This positive
exponent, estimated from experimental time series, is a quantification of the SIC of the Super-ACO FEL.

\section{The route to chaos}

In order to elucidate the emergence of chaos, previously revealed by the positive Lyapunov exponent for example, the modulation amplitude is progressively tuned. The observed dynamical regimes are analysed through their spectrum and their trajectories in phase space.

\subsection{Dynamical regimes increasing the amplitude of the modulation}

\begin{figure}
  \includegraphics[scale=0.35]{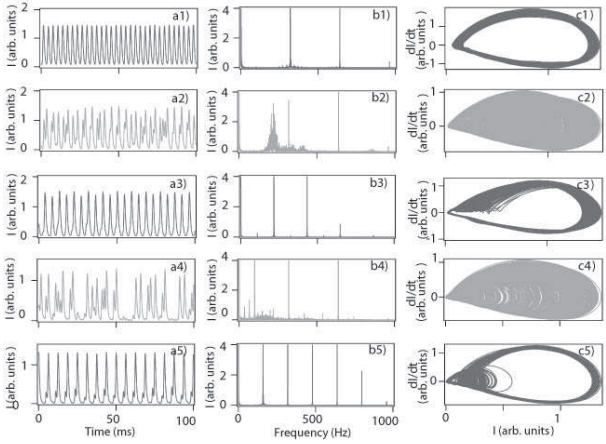}
            \caption{Experimental dynamics observed on the Super-ACO FEL. The laser intensity is displayed in the first column, its spectrum in the second one, and the projection of the trajectory in the reduced phase space in the last column. The so-called derivative coordinates \cite{PhysicaD82Farmer} [$I(t)$, $dI(t)/dt$, $d^{2}I(t)/dt^{2}$,..., $d^{m}I(t)/dt^{m}$] are used to draw a representative picture of the phase space. The lines correspond respectively (from top to bottom) to the cases $a=7,\ 17,\ 28,\ 35$ and $51~fs$. Experimental conditions: $f=320~Hz$, $I_{c}=30~mA$, $G_{0}\approx 1.4~\%$ and $L\approx 0.8~\%$.}\label{sequence-exp-320Hz}
\end{figure}

Figure \ref{sequence-exp-320Hz} shows five experimental dynamical behaviours, representative of those observed during the different
experiments -- there are some uncertainties in the experimental parameters. For a low detuning amplitude $a=7~fs$, a $T$--periodic regime is observed, called $1T$, with $T=1/f$. In this regime, the laser intensity (see figure~\ref{sequence-exp-320Hz}-1) is modulated at the modulation frequency (at $f=320~Hz$), as its spectrum shows.
The projection in the $(I,dI/dt)$ phase space, called hereafter ``reduced phase space'', is close to a closed curve, which means that the system has reached a limit cycle.
At a modulation amplitude of $17~fs$, the laser intensity exhibits an erratic behaviour: The delay between the macropulses and the intensity peaks is now quite irregular (see figure \ref{sequence-exp-320Hz}-2). Inspection of the Fourier space, where a broad spectrum is now observed, confirms this aperiodicity. As the trajectory tends to fill a surface--like area in the reduced phase space, one can conclude on the presence of chaos in the laser.

Then, when the modulation amplitude is increased until $28~fs$, periodicity is back in the system (see figure \ref{sequence-exp-320Hz}a3,b3,c3), and the intensity now oscillates at $2f/3$. Thus the laser response is now at $f/3$ and its harmonics, although the even ones are at a much higher amplitude. The trajectory is a once again a closed curve, and its thickness in the $(I<1,dI/dt>0)$ part attests to the presence of secondary peaks of intensity next to the main one.

When the modulation amplitude is further increased until $35~fs$, the laser dynamics becomes again chaotic. However one must note that although the trajectory fills an area of the reduced phase space similar to the one of the chaotic regime at $17~fs$, the spectrum reveals that different frequencies are dominating the dynamics.

Finally, the modulation amplitude is set at $51~fs$, and the system exhibits a nearly periodic regime, with the main peaks of intensity at a frequency, which is half the one of the $a=7~fs$ case. Once more, the presence of harmonic frequencies can be associated with secondary peaks at a lower intensity.

Thus, the dynamics observed experimentally point out an alternation of chaotic and nearly periodic regimes, which enables the exploration of the {\it route to chaos} in the system, that is the transition from regular dynamics to chaotic ones. Nevertheless, a finer exploration of the bifurcation diagrams remains experimentally problematic, because of a drift in the system parameters. Actually, theoretical and experimental studies on an internal modulated $CO_{2}$ laser have demonstrated that some bifurcations can
disappear because of perturbations present in the system \cite{TheseHennequin}. 


In order to further elucidate the mechanism of emergence of chaos in the system, the theoretical model (\ref{modelEPJD}) is used, since it allows for a finer tuning of the modulation amplitude.

Because of the system parameter drift, a quantitative agreement between the experimental data and the model are hard to achieve. Instead, we choose typical values of parameters of the Super-ACO FEL (that is, $2\%$ and $0.5\%$ respectively for the gain and losses coefficients), and then we observe the succession of regimes which emerge, even if they occur at different modulation amplitudes.

\begin{figure}
  \centering
  \includegraphics[scale=0.5]{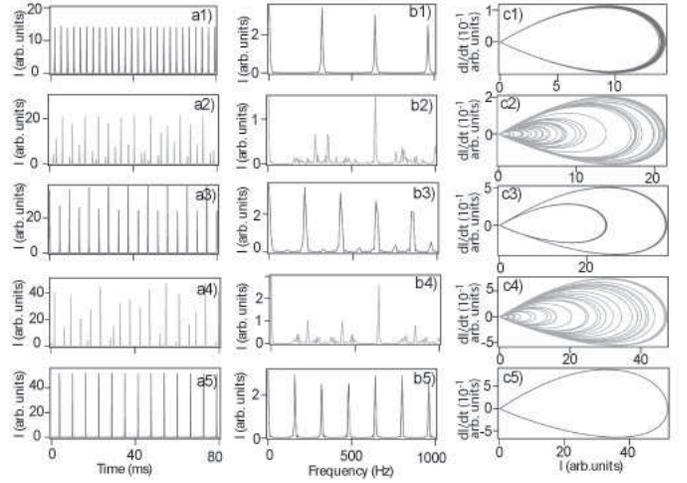}
            \caption{Dynamics of model (\ref{eq:LAS1a},\ref{eq:LAS1b}). The first column shows the laser intensity time series, the second one its spectrum, and in the third one, the trajectory in the reduced phase space is represented. The lines depict, from top to bottom, the regimes at $a=24$, $36$, $108$, $168$ and $216~fs$. The parameters are typical of the Super-ACO FEL: $\tau_{R}=120~ns$, $G_{0}=2~\%$, $L=0.5~\%$, $\sigma_{0}=6.5$x$10^{-4}$, $\sigma_{e}=7.5$x$10^{-4}$ , $\sigma_{\tau}=107~ps$, $\tau_{s}=9~ms$,
$i_{s}=~10^{-9}$, $f=320~Hz$.}\label{sequence-theo-320Hz}
\end{figure}
\begin{figure}[h]
  \centering
  \includegraphics[scale=0.7]{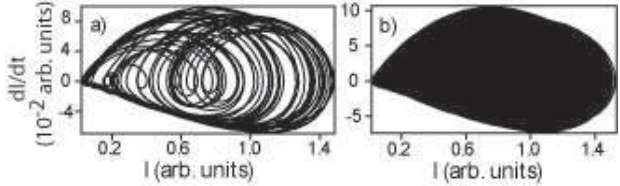}
            \caption{Experimental portion of the system trajectory in the reduced phase space, over a) 100 ms, b)
2000 ms. Experimental conditions: $f=320~Hz$, $a=17~fs$, $I_{c}=30~mA$, $G_{0}=1.4~\%$, $L=0.8~\%$.}\label{portrait-chaos-2500pt}
\end{figure}

Figure \ref{sequence-theo-320Hz} displays the regimes obtained from model (\ref{eq:LAS1a},\ref{eq:LAS1b}). At a low modulation amplitude ($a=24~fs$), the intensity is regularly pulsed, at the frequency $f$ (orbit 1T). When $a$ is increased to $36fs$, the system switches to a chaotic regime, a peak at $2f$ is still present, but the intensity spectrum is quite broad, and the trajectory begins to fill a mussel-like area of the reduced phase space. One has to note that the difference between the experimental and numerical trajectory in this space is essentially due to the measurement characteristic time, which is typically twenty times larger in the former case. Thus, a chaotic trajectory (such as case (c2) of Figs.\ref{sequence-exp-320Hz} and \ref{sequence-theo-320Hz}), which is supposed to be dense in a given area of the reduced phase space appears to fill it very partially because of the finiteness of the measurement time. Moreover, the shorter is this time, the smaller is the fraction of phase space that it seems be filled. This behaviour is illustrated in Fig.\ref{portrait-chaos-2500pt}, where two trajectories, measured over respectively $100ms$ and $2000ms$, are compared.

When $a$ is increased from $36$ to $108$ fs, the intensity starts oscillating at frequency $2f/3$, and its harmonics (orbit 3T): The system is now periodic, as it can be observed in the reduced phase space. The modulation amplitude is then set at $a=168~fs$, a regime in which the intensity is composed of erratic peaks. the spectrum is broad, and on a time interval of $100~ms$, the trajectory starts filling a surface of the reduced phase space,which is then a signature of chaos. Finally, when $a$ reaches $216~fs$, the intensity is pulsed periodically at $f/2$, and the system seems to have a periodic behaviour.

\begin{figure}
  \centering
  \includegraphics[scale=0.7]{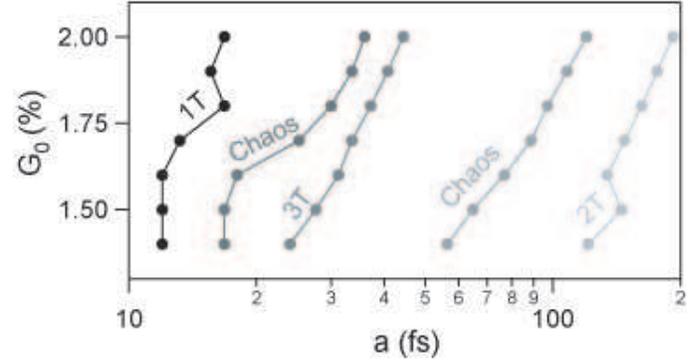}
            \caption{Amplitude for which the transition of the dynamical regimes occurs as a function of the initial gain with fix losses. Same parameters as for figure \ref{sequence-theo-320Hz}.}\label{diag_stab}
\end{figure}

Thus, it was observed the succession of the 1T-chaos-3T-chaos-2T regimes with the model (\ref{eq:LAS1a},\ref{eq:LAS1b}), which gives a {\it qualitative} agreement between the model and the experiments. Once more, the {\it quantitative} one remains problematic because of the system parameter drift in particular due to the electron bunches lifetime (which affects the number of electrons in the bunch and thus the gain). Figure \ref{diag_stab} shows that for different gain values, the succession of the regimes remains the same even if the amplitude for each transition changes with the gain. The model (\ref{eq:LAS1a},\ref{eq:LAS1b}) reproduces well the alternation of periodic and chaotic regimes. 

%

\subsection{Period-doubling bifurcation}
\begin{figure}[h]
  \centering
  \includegraphics[scale=0.9]{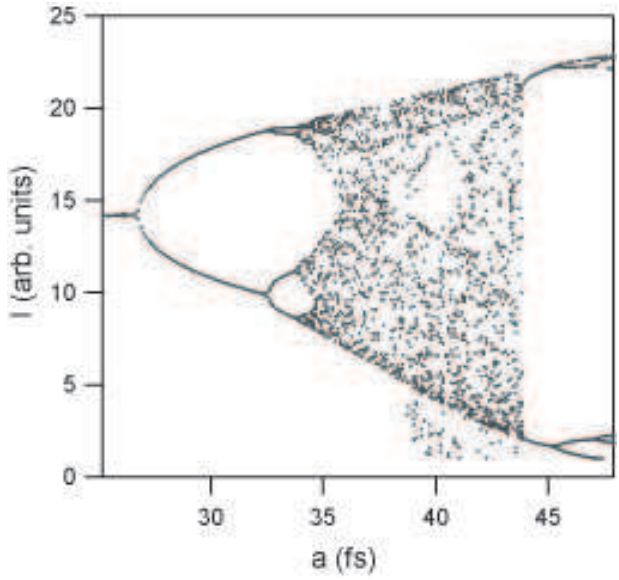}
            \caption{Simulated bifurcation diagram of the SRFEL: The intensity peaks are drawn versus the modulation amplitude. The cascade of periodic regimes eventually turns into chaos, before returning to periodicity. Same parameters as for figure \ref{sequence-theo-320Hz}.\label{fig:RouteToChaos}}
\end{figure}

\begin{figure}[h]
  \centering
  \includegraphics[scale=0.78]{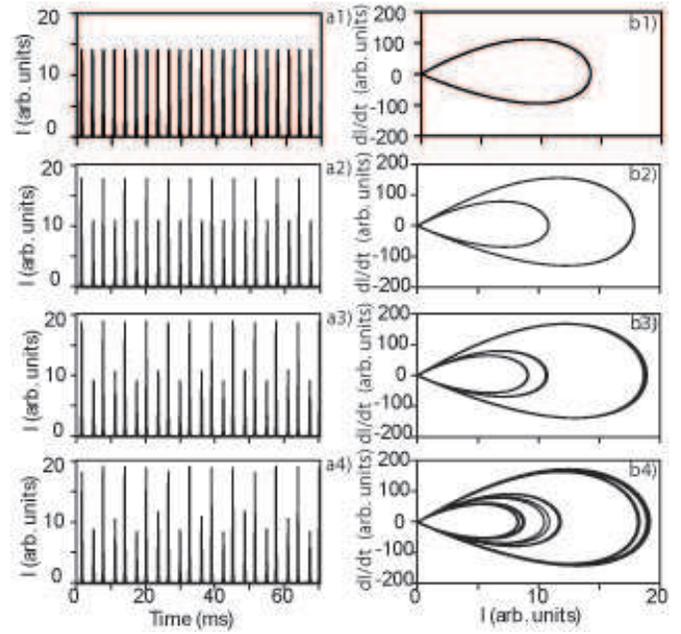}
            \caption{Dynamics of model (\ref{eq:LAS1a},\ref{eq:LAS1b}) for a fine increase of the modulation amplitude. In the left column is displayed the intensity evolution, in the right one, the trajectory in the reduced phase space. The three lines correspond, from top to bottom, to $a=30$, $33$ and $34.2~fs$. Same parameters as for figure \ref{sequence-theo-320Hz}.}\label{Feigenbaum-theo-320Hz}
\end{figure}
In order to elucidate the mechanism of emergence of chaos at a finer scale than the one allowed experimentally (not speaking of the parameters drift), the modulation amplitude is slowly increased in model (\ref{eq:LAS1a},\ref{eq:LAS1b}), using the previous set of parameters. Fig.\ref{Feigenbaum-theo-320Hz} displays the evolution of intensity and of the system trajectory in the reduced phase space when $a$ is increased. At $a=30~fs$, the orbit is 2T. Indeed, although the intensity peaks still occur every $2/f$, their peak value is now modulated at half the modulation frequency. Then, an increase of $3~fs$ leads to a doubling of the period of the dynamics, that is an obit 4T: The intensity maxima are now modulated at $f/4$. One more increase of $1.2~fs$ leads to an extra demultiplication of the orbit branches. And as seen previously, a little beyond that value (at $a=36~fs$), the system dynamics becomes chaotic (see Fig.\ref{sequence-theo-320Hz}). This period-doubling cascade is illustrated in Fig.\ref{fig:RouteToChaos} where the successive peaks of intensity are drawn versus the modulation amplitude. The doubling of the period of the regimes occurs at $a_{1T\rightarrow 2T}\approx 27 fs$, then at $a_{2T\rightarrow 4T}\approx 33 fs$, before the system enters a chaotic regime around $a_{4T\rightarrow C}\approx 34 fs$. The return to a periodic $2T$ regime is observed around $a_{C\rightarrow 2T}\approx 44 fs$. New bifurcations appear at higher values of $a$, which are not studied here. This study at constant gain/losses parameters demonstrates that the Super-ACO FEL is entering a chaotic regime via a period-doubling cascade, a traditional ``route to chaos'' (see fig. \ref{fig:RouteToChaos}).

For a higher ratio gain/losses (typically 3 times higher) than the experimental one, larger modulation amplitude step is required to go from orbit nT to 2nT. Since the step to go from 2T to 4T in the model is $3~fs$, the experimental observation of the doubling-period cascade is estimated around $1 fs$, which is less than the step applied for the modulation amplitude ($1.2~fs$). So that only the alternation of chaotic and regular regimes is observed experimentally.
The bifurcation diagram of the Super-ACO FEL can be compared to the bifurcation diagram of the logistic map. The
first transition to chaos occurs through a period doubling cascade. Then other routes as intermittence should be observed, leading to periodic windows.


\subsection{Intermittence}

\begin{figure}[h]
  \centering
  \includegraphics[scale=0.8]{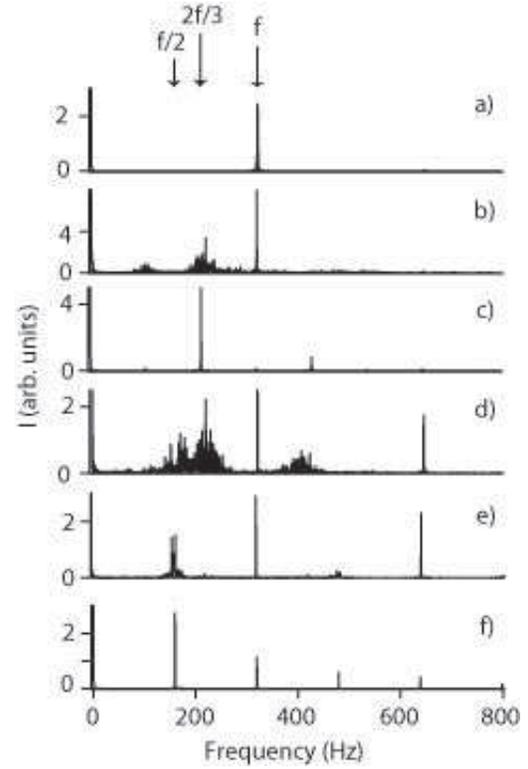}
            \caption{Spectrum of the laser intensity for different amplitudes of the modulation.
            a) 1T behavior: a =7 fs, b) 3C behavior: a= 17 fs, c) 3T behavior: a= 28 fs, d) 3C behavior: a= 35 fs, e)
            2C behavior: a= 50 fs, f) 2T behavior: a= 51 fs. Experimental conditions: $I_{c}=30~mA$, $G_{0}\approx 1.4~\%$,
            $L\approx 0.8~\%$, $f=320~Hz$.}\label{evolution-chaos}
\end{figure}

A first manifestation of intermittence is the difference between the two experimental chaotic behaviors at $a= 17 fs$ and $a=35 fs$ (see figure\ref{sequence-exp-320Hz}). The spectrum of the first chaotic motion
has a dominant frequency at 2f/3 (see figure \ref{evolution-chaos}b). This is called hereafter 3C: C for
chaos and 3 for 3T. As a consequence, one can expect that a 3T-periodic orbit exists for close values of parameters, as it is effectively
observed for $a=28fs$ (see figure \ref{evolution-chaos}c). By increasing the amplitude of the modulation, the spectrum blows up
again, with a dominant frequency at f/3 (see figure \ref{evolution-chaos}d). This 3C evolves
towards a chaotic regime with a dominant frequency at f/2, called 2C (see figure \ref{evolution-chaos}e). The laser then becomes 2T periodic (see figure \ref{evolution-chaos}f). A similar evolution
of the attractor has been also observed with a modulated $CO_{2}$ laser \cite{TheseHennequin}. This indicates that chaos and periodic dynamics may occur through intermittence in a storage-ring FEL.
\begin{figure}[h]
  \centering
  \includegraphics[scale=0.6]{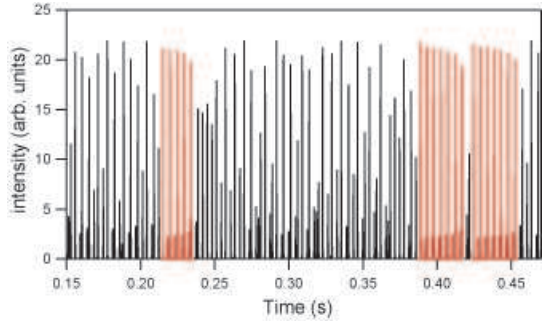}
            \caption{Simulated laser intensity versus time with intermittence phenomena highlight in red. a=43.74fs.}\label{ex_interm}
\end{figure} 
\begin{figure}[h]
  \centering
  \includegraphics[scale=0.6]{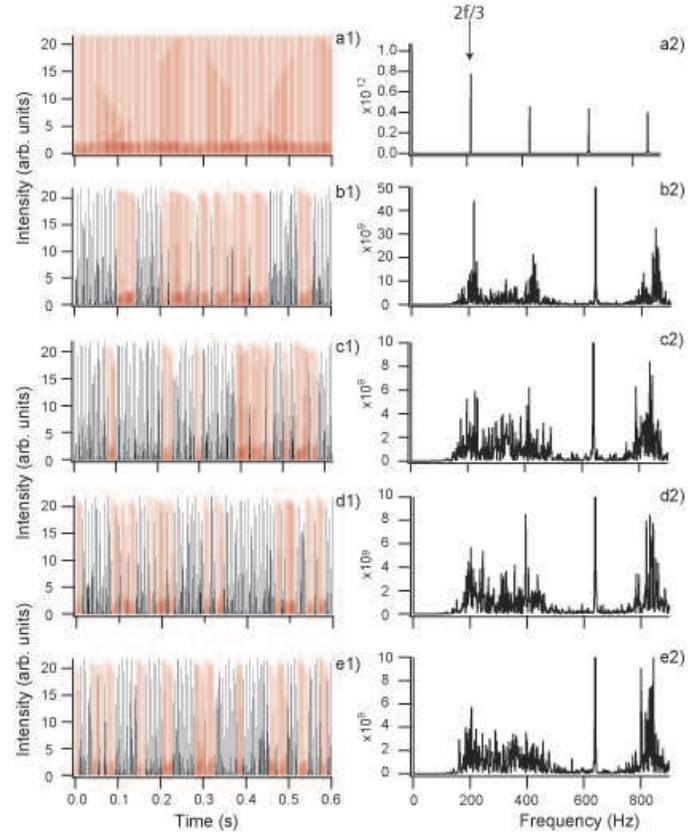}
            \caption{Evolution of the intermittence with the modulation amplitude and its consequence on the spectrum.1) Time series of the laser intensity and its 2) power spectrum. a) a=44.04fs, b) a=43.8 fs, c)a=43.74fs, d) a=43.68fs, e) a=43.62 fs. From e) to a), the periodic outburst in the chaotic time series increases. Similarly, the spectrum shows a growth of the peak frequency at 2f/3 with f=320 Hz.}\label{interm-simus}
\end{figure}
A more detailed picture of the phenomenon can be found thanks to the model. In the simulated time series, burst of periodic signals appear in the chaotic signal (see fig. \ref{ex_interm}). The time between successive bursts tends to increase when getting closer to the periodic regime. The spectrum corresponding to a chaotic time series also exhibits a dominant frequency given by the periodic parts (see fig \ref{interm-simus}), whose peak, at 2f/3, increases with the amplitude of the modulation. When the intermittence is well established, the system switches to a periodic behavior whose frequency corresponds to the dominant peak in the adjacent chaotic windows.

\subsection{Symmetry of the system}
\begin{figure}[h]
  \centering
  \includegraphics[scale=0.5]{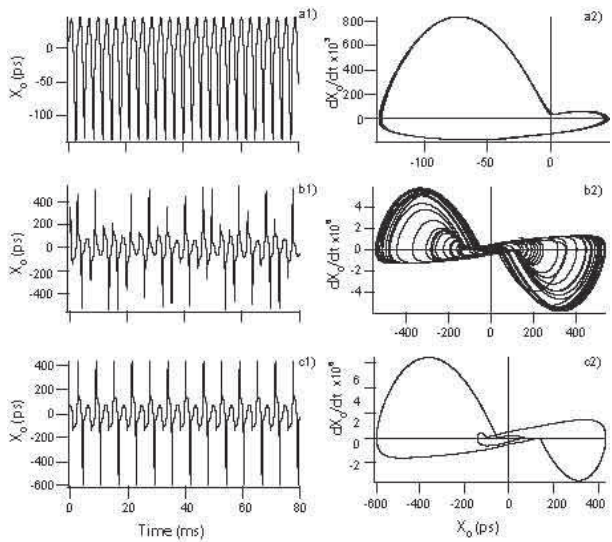}
            \caption{Numerical evolution of the $\emph{center~of~mass}$ of the FEL pulse for a modulation frequency f= 320 Hz. 1) Time series and 2) reconstructed phase space of the moment of order two of the FEL pulse distribution. a) 1T motion : a= 24 fs, b) Chaotic motion : a= 168 fs, c) 2T motion : a= 216 fs. Numerical parameters of figure \ref{sequence-theo-320Hz}. The FEL pulse center~of~mass is retrieved and analysed, through the first order momentum of the distribution.}\label{pos-num-320}
\end{figure}

A manifestation of symmetry can be found in the time series given by the position of the $\emph{center~of~mass}$ of the light pulse called in the following as position. Figure \ref{pos-num-320} shows the numerical results on the evolution of the laser pulse position. In all cases (periodic and chaotic) the evolution is periodically pulsed at the modulation frequency. The evolution of the amplitude of the position is well described by the projected phase space reconstruction from the derivation coordinates. For periodic cases, the phase space shows an asymmetric limit cycle, whereas a symmetric attractor is revealed for the chaotic regimes. In the system, there is a natural symmetry due to the fact that the modulation is symmetric compared to the zero temporal detuning condition, leading to a pulse position which is naturally either positive or negative. These various symmetries can be used to understand the chaotic dynamical behavior, which opens new possibilities for a simplification of the model. Note that the possibility for chaos of exhibiting symmetries is still a debating topic and it has been the subject of different studies \cite{PRE95Letellier,PhysicaD02BenTal}.

\section{Conclusion}

In this paper, we have studied a particular
case of electron beam instabilities by applying an external modulation periodically modulated near the resonant frequency of the storage ring FEL 
system. It lead us to characterize the chaotic response of the Super-ACO FEL to such an electron beam instability.

By calculating the embedding dimension of the time series, we demonstrate that a dimension three is
sufficient to describe accurately the system phase space, including its nonlinear features, in spite of the infinite dimension of the model. The chaos
is characterized by a positive Lyapunov exponent, which reveals the sensitivity of the time series to the initial conditions in the case of the Super-ACO FEL. The experimental data, combined with the finer analysis of the model, leads to the conclusion that the first transition to chaos is a period doubling cascade and the successive transition may occur by intermittence. The intermittence has been revealed experimentally by the evolution of the spectrum, where peaks could be observed as a sign of periodic orbits at close values of parameters. This signature of intermittence in the spectrum has been confirmed numerically. This analysis of the chaos reveals that a non conventional laser like the FEL has a bifurcation diagram very similar to the one of the logistic map. As a consequence, stabilisations of this kind of FEL instabilities can be envisaged and inspired for developing feedback systems, as it was already developed for table top lasers for example.



\section{Acknowledgments}
The authors acknowledge J. Polian, F. Ribeiro and R. Lopes, B. Rieul, T. Guillou, the research group (GDR) of nonlinear physics, S. Bielawski, C. Letellier and R. Carr for our fruitful discussions.

\bibliographystyle{epj}
\bibliography{stab}
\end{document}